\documentclass[intlimits,twoside,a4paper]{article}
\usepackage{wrapfig}
\usepackage{graphicx}
\usepackage[T2A]{fontenc}
\usepackage{amssymb}
\usepackage[cp1251]{inputenc}

\usepackage[]{cmpj2}
\articletype{Rapid communication}

\newcommand{\be}{\begin{equation}}
\newcommand{\ee}{\end{equation}}
\newcommand{\bea}{\begin{eqnarray}}
\newcommand{\eea}{\end{eqnarray}}

\newcommand{\dtwo}[2]{\frac{\partial^2 #1}{\partial #2^2}}
\newcommand{\done}[2]{\frac{\partial #1}{\partial #2}}
\newcommand{\eps}{\varepsilon}

\title[Transverse Piezoelectric Resonance in KH$_2$PO$_4$ Type
Crystals]{Transverse Piezoelectric Resonance in KH$_2$PO$_4$ Type
Crystals}
\author[R.R. Levitskii,
    I.R. Zachek, A.P. Moina, A.S. Vdovych]{R.R. Levitskii\refaddr{label1},
    I.R. Zachek\refaddr{label2}, A.P. Moina\refaddr{label1}, A.S. Vdovych\refaddr{label1}}
\addresses{
\addr{label1} Institute for Condensed Matter Physics of the
National Academy of Sciences of Ukraine, 1 Svientsitskii str,
79011 Lviv, Ukraine \addr{label2} Lviv Polytechnic National
University, 12 Bandera Str., 79013 Lviv, Ukraine }
\begin{document}

\maketitle

\begin{abstract}
Within the framework of proton model with taking into account the
piezoelectric interaction with the shear strain $\varepsilon_4$, a
dynamic dielectric response of KH$_2$PO$_4$ family crystals to the
electric field perpendicular to the axis of spontaneous
polarizaiton is considered. Piezoelectric resonance frequencies of
rectangular thin plates of the crystals cut in the (100) plane
(0$^\circ$ X-cut) are calculated.
\pacs 77.22.Ch, 77.22.Gm,  77.84.Fa, 77.65.Fs

\end{abstract}

\section{Introduction}

In our previous paper \cite{new-kdp} we explored the dynamic
dielectric response of square thin plates, cut from the
KH$_2$PO$_4$ family ferroelectric and antiferroelectric crystals
in the planes (001), to the a.c. electric field applied along the
axis [001]. This is the axis of a spontaneous polarization in the
KH$_2$PO$_4$ type ferroelectrics, and we call this field
longitudinal.

Using the modification of the proton ordering model \cite{our!!}
that includes the piezoelectric coupling with the shear strain
$\eps_6$, within the framework of the Glauber approach \cite{61x}
and the four-particle cluster approximation, we obtained
expressions for the dynamic dielectric permittivity of the
crystals, which took into account the dynamics of the shear strain
$\eps_6$. The  found expressions for the  resonant frequencies of
the longitudinal dielectric permittivity are in a good agreement
with experiment.

Useful information can be obtained also by investigating the
transverse dielectric response of ferroelectric crystals, and
especially  the proton glass systems of the
Rb$_{1-x}$(NH$_4)_x$PO$_4$ type, when the applied electric field
is perpendicular to the axis of spontaneous polarization. For the
crystals of the KH$_2$PO$_4$ family that would be the axis [100].
The field $E_1$ induces the shear strain $\eps_4$ via the
piezoelectric coefficient $d_{14}$, and tThe corresponding
dielectric permittivity exhibits a piezoelectric resonant
dispersion. Frequencies of these resonances will be determined in
the present paper. In our calculations we shall use the models
presented in \cite{ntsh,preprint-kdp,preprint-adp}, which take
into account the piezoelectric coupling to the shear strain
$\eps_4$.

\section{Transverse dynamic permittivity of KH$_{2}$PO$_{4}$ type
crystals}

We shall consider shear mode vibrations of a thin $L_y \times L_z$
rectangular plate of a KH$_2$PO$_4$ crystal, cut in the (100)
plane, with the edges along [010] and [001] ($0^\circ$ X-cut). The
vibrations are induced by time-dependent electric field
$E_{1t}=E_1e^{i\omega t}$.

Dynamics of pseudospin subsystem will be considered in the spirit
of the stochastic Glauber model \cite{61x}, using the
four-particle cluster approximation.  The  system of equations for
the time-dependent deuteron (pseudospin) distribution functions is
  \be
  \label{3.1}
  - \alpha \frac{d}{dt} \langle \prod_f \sigma_{qf} \rangle =
  \sum\limits_{f'} \langle \prod_f \sigma_{qf} \left[ 1 -
  \sigma_{qf'} \tanh \frac12 \beta {\mbox{\boldmath$\varepsilon$}}_{qf'}(t)
  \right] \rangle ,
  \ee
where ${\mbox{\boldmath$\varepsilon$}}_{qf'}(t)$ is the local
field acting on the  $f'$th
 deuteron in the $q$th cell, which can be found from the system Hamiltonian \cite{ntsh,preprint-kdp,preprint-adp}; $\alpha$ is the parameter setting the time scale of
 the dynamic processes in the pseudospin subsystem.

Dynamics of the deformational processes is described using
classical Newtonian equations of motion of an elementary volume,
which for the relevant to our system displacements $u_1$ and $u_2$
($\varepsilon_4 =  \frac{\partial u_2}{\partial
 z} + \frac{\partial u_3}{\partial y}$) read
\begin{equation}
\label{3.6} \rho\dtwo{u_2}{t}=\done{\sigma_{4}}{z},\quad
\rho\dtwo{u_3}t=\done{\sigma_{4}}{y}.
\end{equation}
Here $\rho$ is the crystal density,  $\sigma_{4}$ is the
mechanical shear stress,  which, being the function of
$\eta^{(1)}$, $E_1$, and $\varepsilon_4$, is found from the
constitutive equations derived in
 \cite{old}.

Following the usually used procedure
\cite{new-kdp,old,old-adp,ntsh}, at small deviations from the
equilibrium, we derive the following equations for the
displacements
 \begin{eqnarray}
 \label{3.7}
 && \frac{\partial^2 u_{1}}{\partial y^2} + k_{4}^2u_{1} = 0,
 ~~~ \frac{\partial^2 u_{2}}{\partial x^2} + k_{4}^2u_{2} = 0,
 \end{eqnarray}
where $k_{4}$ is the wavevector
 \be
 \label{7.14}
 k_{4} = \frac{\omega
 \sqrt{\rho}}{ \sqrt{c_{44}^E(\alpha\omega)} }.
  \ee
Expressions for the frequency dependent elastic constants
$c_{44}^E(\alpha \omega)$ for the KH$_2$PO$_4$ type ferroelectrics
and NH$_4$H$_2$PO$_4$ type antiferroelectrics are presented in
\cite{ntsh,preprint-kdp,preprint-adp}.

  Differentiating the first and second equations of (\ref{3.7}) with
  respect to $z$ and $y$, correspondingly, and adding the two obtained equations, we
  arrive at the single equation for the strain $\eps_4$
\begin{eqnarray}
&& \label{singlesystem}
\dtwo{\varepsilon_{4}(y,z)}{y}+\dtwo{\varepsilon_{4}(y,z)}{z}+k_4^2\varepsilon_{4}(y,z)=0.
\end{eqnarray}
Boundary conditions for $\varepsilon_{4}(y,z)$ follow from the
assumption that the crystal  is traction free at its edges (at
$y=0$, $y=L_y$, $z=0$, $z=L_z$, to be denoted as $\Sigma$)
\begin{equation}
\label{boundary1} \sigma_4|_\Sigma=0.
\end{equation}
Solution of (\ref{7.14}) with the boundary conditions
(\ref{boundary1}) is
\begin{equation}
\label{strain}
\eps_{4}(y,z)=\eps_{40}
+ \eps_{40}\sum_{k,l=0}^\infty
\frac{16}{(2k+1)(2l+1)\pi^2}\frac{\omega^2}{(\omega_{kl}^0)^2-\omega^2}\sin\frac{\pi(2k+1)y}{L_y}\sin\frac{\pi(2l+1)z}{L_z},
\end{equation}
where
\[
\eps_{40}=\frac{e_{14}(\alpha\omega)}{c_{44}^E(\alpha
\omega)}E_{1t};
\]
 $\omega^0_{kl}$ is given by
\begin{equation}
\label{res4} \omega_{kl}^0=\sqrt{\frac{
c_{44}^E(\omega^0_{kl})\pi^2}{\rho}\left[\frac{(2k+1)^2}{L_y^2}+\frac{(2l+1)^2}{L_z^2}\right]},
\end{equation}
$e_{14}(\alpha\omega)$ is the piezoelectric coefficient, the
expressions for which for the KH$_2$PO$_4$  and NH$_4$H$_2$PO$_4$
type crystals, have been derived in
\cite{ntsh,preprint-kdp,preprint-adp}.

The transverse dynamic dielectric susceptibility of a free crystal
has been obtained in the following form \cite{ntsh}
 \begin{equation}
 \label{dyn_susc}
\chi_{11}^\sigma(\omega)=\chi_{11}^{\eps}(\alpha\omega) +
R_4(\omega)\frac{e_{14}^2(\alpha\omega)}{c_{44}^E(\alpha\omega)},
\end{equation}
 where
 \begin{equation}
 \label{r6}
 {R_4(\omega)}=1+ \sum_{k,l=0}^\infty
\frac{64}{(2k+1)^2(2l+1)^2\pi^4}\frac{\omega^2}{(\omega_{kl}^0)^2-\omega^2};
 \end{equation}
 $ \chi_{11}^\eps(\alpha\omega)$
is the dynamic dielectric susceptibility of a clamped crystal. The
corresponding expressions can be found in
\cite{ntsh,preprint-kdp,preprint-adp}.

The static and the high frequency limits  of (\ref{dyn_susc}) are
the static susceptibility of a free crystal \cite{our!!} and the
dynamic susceptibility of a mechanically clamped crystal,
exhibiting relaxational dispersion in the microwave region. Thus,
eq. (\ref{dyn_susc}) explicitly describes the effect of crystal
clamping by high-frequency electric field.

 In the
intermediate frequency region, the susceptibility has a resonance
dispersion with numerous peaks ar frequencies where ${\rm
Re}[R_4(\omega)]\to\infty$. Frequency variation of
$c_{44}^E(\alpha\omega)$ is perceptible only in the  region of the
microwave dispersion of the dielectric susceptibility. Below this
region it is practically frequency independent and coincides with
the static elastic constant $c_{44}^E$. Since the resonance
frequencies are expected to be in the $10^4-10^7$~Hz range,
depending on temperature and sample dimensions,  the equation for
the resonance frequencies (\ref{res4}) is reduced to an explicit
expression by putting in it $c_{44}^E(\alpha\omega)\to c_{44}^E$.

\begin{figure}[hbt]
\centerline{\includegraphics[width=0.7\textwidth]{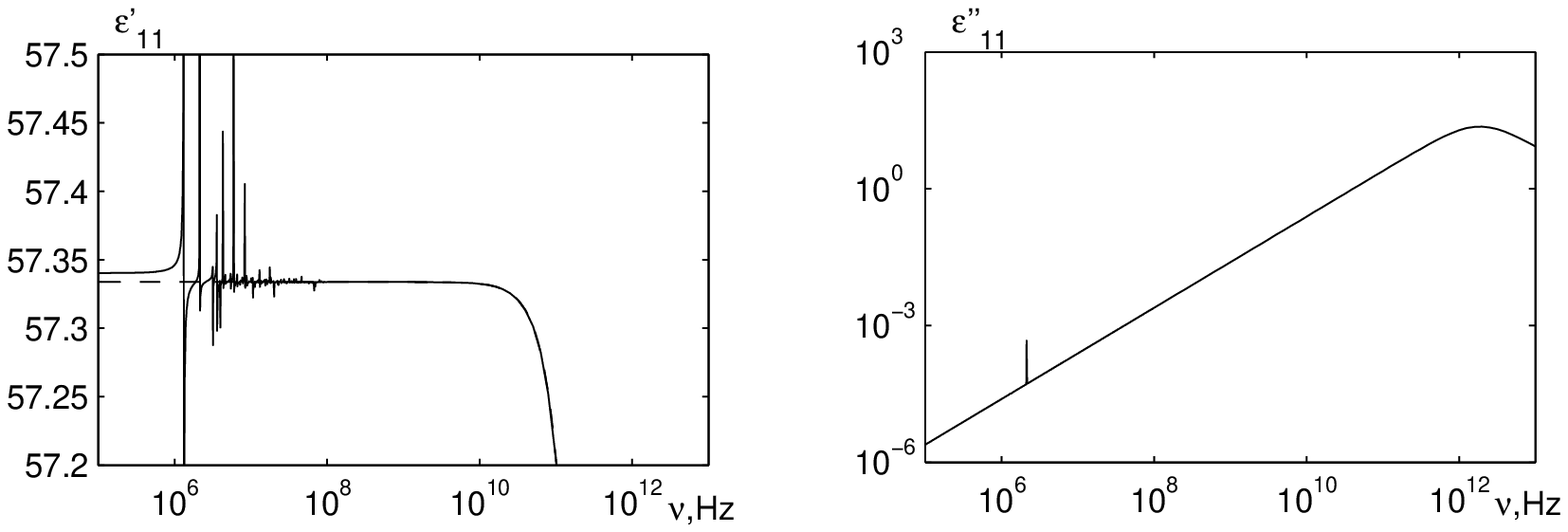}}
\centerline{\includegraphics[width=0.7\textwidth]{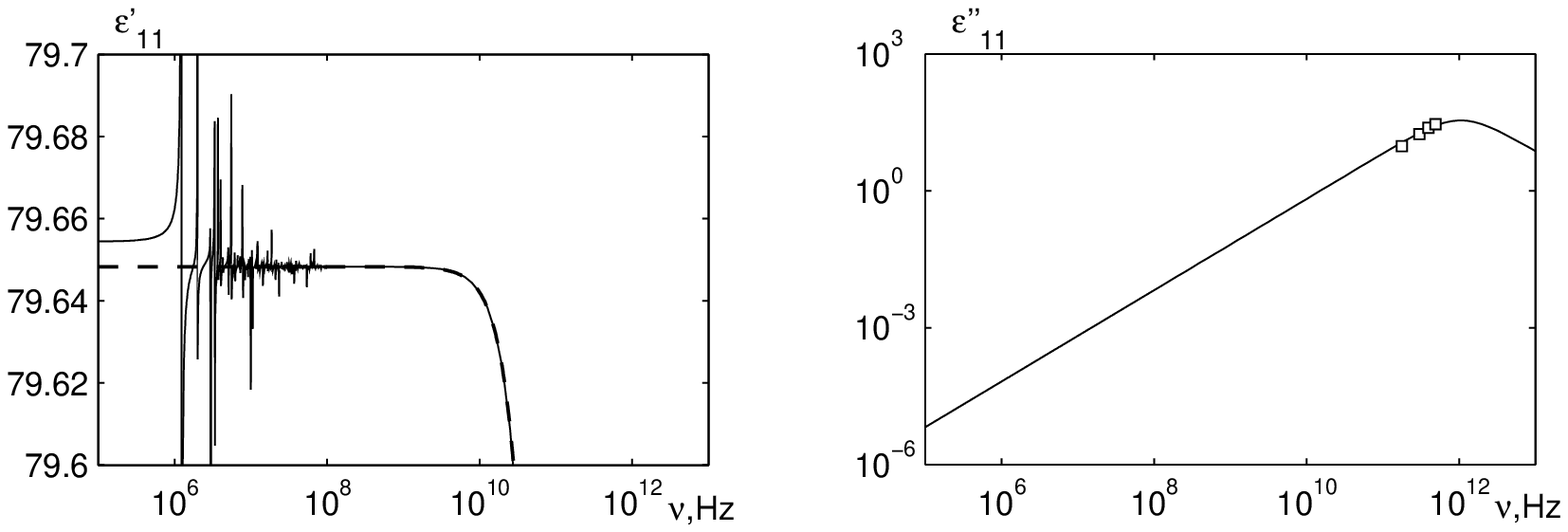}}
\caption{The frequency dependences of the transverse dynamic
dielectric permittivity for KH$_2$PO$_4$ (top) and
NH$_4$H$_2$PO$_4$ (bottom) crystals at 127.5~K ($T=T_{c}+5$~K) and
183~K ($T=T_{\rm N}+35$~K), respectively. $L_y=2$~mm, $L_z=1$~mm.
Lines: a theory; $\square$: experimental points of \cite{465x}.
Solid and dashed line: the permittivity of a free and clamped
crystal, respectively.}\label{fig1}
\end{figure}

In figure~\ref{fig1} we plotted the frequency dependences of the
transverse dynamic dielectric permittivity for KH$_2$PO$_4$ and
NH$_4$H$_2$PO$_4$ crystals. The used values of the  model
parameters can be found in \cite{ntsh,preprint-kdp,preprint-adp}.
For the other ferroelectric and antiferroelectric crystals of the
KH$_2$PO$_4$ family, obtained by isomorphic replacement of K and P
ions, these dependences are totally analogous. Evolution of the
permittivity with increasing frequency from the free crystal value
through the piezoelectric resonances to the clamped crystal value
and then to the relaxational dispersion in the microwave region is
observed for all crystals of the family.

\begin{figure}[hbt]
\centerline{\includegraphics[width=0.35\textwidth]{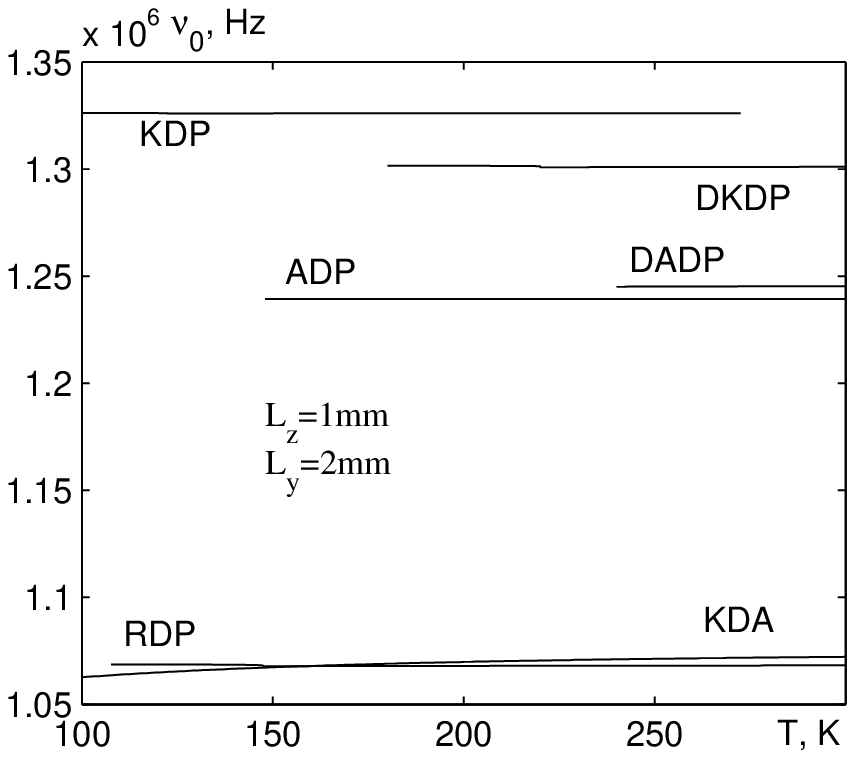}}
\caption{The temperature dependences of the first resonant
frequency of the transverse dynamic dielectric permittivity for
different crystals of the KH$_2$PO$_4$ familly. $L_y=2$~mm,
$L_z=1$~mm. }\label{fig2}
\end{figure}

Since the elastic constant $c_{44}^E$ shows no significant
anomalies at the transition point or temperature variation in
these crystals \cite{Mason,Mason2}, the first resonant frequency
of the transverse permittivity, according to (\ref{res4}) at
$k=l=0$, is temperature independent as well. This is illustrated
in fig.~\ref{fig2}, where these frequencies are given for
different crystals of the KH$_2$PO$_4$ family.

\section{Conclusions}

Within the proton ordering model with taking into account the
shear strain $\varepsilon_4$ we explored a dynamic response of
ferroelectric and antiferroelectric crystals of the  KH$_2$PO$_4$
family to a transverse external harmonic electric field $E_1$.
Corrected expressions for the piezoelectric resonance frequencies
of simply supported rectangular 0$^\circ$ X-cuts of these crystals
are obtained. The ultimate goal of the present studies  will be to
generalize the obtained expression for the dynamic permittivity to
the case of the Rb$_{1-x}$(NH$_4)_x$PO$_4$ type  proton glasses,
in order to explore their dynamic dielectric response.

\section*{Acknowledgement}
The authors acknowledge support from the State Foundation for
Fundamental Studies of Ukraine, Project  ``Electromechanical
nonlinearity of mixed ferro-antiferroelectric crystals of
dihydrogen phosphate family'' No F53.2/070.


%
\sloppy

\end{document}